# Mechanism for the Anomalous Roll Moment in Wraparound Fin Projectiles

N. Ananthkrishnan[1]

*Yanxiki Tech, 152 Clover Parkview, Koregaon Park, Pune 411001, India*

**The occurrence of asymmetric rolling moments in wraparound fin (WAF) projectiles, even at zero free stream angle of attack, and their potential to induce roll reversal and resonant instabilities has long been recognized. The induced rolling moment is seen to flip sign depending on the WAF configuration (with concave surface leading into the roll or vice versa), and between subsonic and supersonic Mach numbers. In this work, we propose a simple mechanism that can explain the anomalous rolling behavior of WAF projectiles in the subsonic regime. By modeling the curved fin by discrete span-wise linear segments, the wraparound fin can be approximated as a wing with a deflected winglet, and the experimental data available in the literature for such wing configurations can be applied to the curved fin of WAF projectiles. The lifting characteristic of the curved fin can be seen to differ depending on whether the wingtip appears to be deflected up or down relative to the local flow incidence angle. Based on this differential lift between a pair of oppositely-placed curved fins on a WAF rocket, it becomes possible to explain the observed sign of the induced rolling moment as well as its dependence on the free stream angle of attack for small values of the angle of attack. The physical mechanism for the differential lift, and hence the induced rolling moment, can be traced to the span-wise location and the liftoff height of the tip vortex from the wraparound fin tips.**

## Nomenclature

$C_l$    =   rolling moment coefficient

$D$    =   body diameter; reference length

$M$    =   Mach number

---

[1] Technical Consultant, Associate Fellow AIAA; akn.korea.19@gmail.com, akn@aero.iitb.ac.in

| | | |
|---|---|---|
| $\alpha$ | = | local angle of attack |
| $\delta$ | = | wingtip segment deflection angle relative to wing/fin main segment |

*Subscripts*

| | | |
|---|---|---|
| 0 | = | value at zero free stream angle of attack |
| $\infty$ | = | free stream condition |

## I. Introduction

Wraparound fins (WAF) are commonly used in tube-launched rockets due to their convenience in packaging. Upon launch, the fins snap open and are then held in place, usually by a spring-loaded mechanism. Where longitudinal aerodynamics are concerned, wraparound fins behave very similarly to their equivalent planar fin (PF) counterparts. However, it has long been known that wraparound fins show anomalous, even erratic, behavior in roll that is not seen in their planar-fin equivalents [1,2]. In particular, the induced rolling moment (usually represented by the static rolling moment coefficient, which is a function of angle of attack and Mach number) in case of wraparound fins appears to flip sign between subsonic and supersonic Mach numbers. One consequence of this flip is the possibility of roll reversal in flight. An example in Ref. [3] shows the projectile initially spin in one sense, then, after some distance downrange, reverse the direction of roll to spin in the opposite sense. The projectile in this example is further seen to temporarily settle down to a roll rate quite close to the pitch-yaw frequency resulting in roll-yaw resonance, followed by a sharp jump in the spin to a much higher value. The possibility of roll-yaw resonance for finned projectiles leading to roll lock-in and resultant large-amplitude yawing motion is well known [4,5]. Even a transient passage though resonance, as in the case of the example in Ref. [3], is sufficient to induce severe yawing motion that can lead to increased dispersion, and, in the worst case, to instability and loss of the rocket [6]. The asymmetry in the rolling moment characteristics can lead to stability boundaries that are biased in one sense of the roll rate, so much so that the zero-spin state can actually be unstable [7]. Moreover, the nature of the asymmetric rolling moment appears to depend on the manner in which the WAF rocket is spun; that is, whether it rolls with the concave side leading or with the convex side leading. The sign of the induced roll moment is different in the two cases, markedly so in subsonic flight. Not surprisingly, a similar asymmetric feature is noted in the roll damping derivative, which happens to depend on the sign of the roll rate [8,9].

In order to obtain a consistently predictable behavior of the dynamics of a WAF rocket in flight, it is necessary to get a handle on the underlying mechanism responsible for the asymmetric rolling moment characteristics, and there has been much effort over the years in this direction. Many experimental and computational studies have focused on the flow phenomena in the vicinity of the curved fins at supersonic Mach numbers [10,11,12]. While agreement between the numerical results and experimental data was generally achieved, no clear fluid mechanic phenomenon could be identified as causing the observed variation in the induced rolling moment. The influence of specific fin design parameters on the nature of the induced rolling moment has been studied as well [13,14]. On the other hand, Ref. [15] considered the possibility of the asymmetric rolling moment in subsonic flow being unrelated to the fin geometry and being caused instead by the presence of the base flow. Still others have attempted to explain the anomalous rolling behavior in terms of the coupling of the rolling motion with the pitch-yaw dynamics, especially the nonlinear side moments [16,17]. However, as summarized in a recent paper [18], no convincing physical explanation for the errant rolling behavior of wraparound fin projectiles has emerged thus far.

In this paper, we put forward a simple mechanism to explain, in a qualitative manner, the induced rolling moment characteristics of wraparound fins at subsonic Mach numbers, as observed from the experimental data reported in Ref. [18]. Specifically, we approximate the curved fin by a discrete number (generally, two) of span-wise linear segments, and then use the available experimental data for such wing configurations in the literature to deduce the nature of the induced rolling moment for different wraparound fin configurations. The mechanism is able to explain the sign of the induced rolling moment in the subsonic regime for both WAF configurations — with the concave side leading into the roll and vice versa. Using this mechanism, it is possible to understand the near-constant value of the induced rolling moment for small angles of attack. The span-wise location and the liftoff height of the tip vortex from the wraparound fin tips appears to be the leading cause of the anomalous rolling moment characteristics in the subsonic regime.

## II. Experimental data for induced rolling moment

First of all, we reproduce the data for the induced rolling moment from Ref. [18] and point out the key features therein. Figure 1(a) shows the variation of the induced rolling moment coefficient with Mach number for a WAF configuration with the concave side leading into the roll. The coefficient (with zero fin cant angle) is positive for all subsonic Mach number values, starts falling around Mach 1.0 crossing the zero line near Mach 1.4, and is then negative for higher supersonic Mach numbers. A similar plot for the case where the convex WAF side leads into the roll is shown in Fig. 1(b), where the subsonic values of the induced rolling moment coefficient (with zero fin cant angle) are negative, with the magnitude decreasing with increasing Mach number. However, in this case, the rolling moment coefficient does not change sign, though it may possibly do so at a Mach number greater than 2.0. Both sets of data are for otherwise identical projectile geometry with body length *L=16.46D*, fin span of *2.10D*, fin chord of *1.68D*, and fin leading edge sweepback of *20* deg, where *D* is the body diameter.

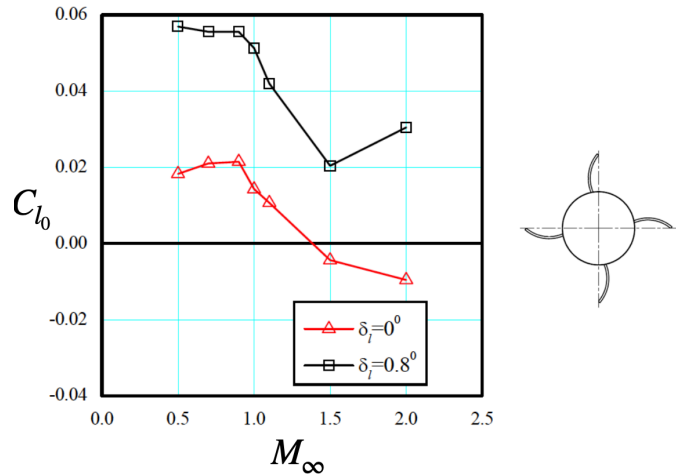

(a) concave side leading into the roll

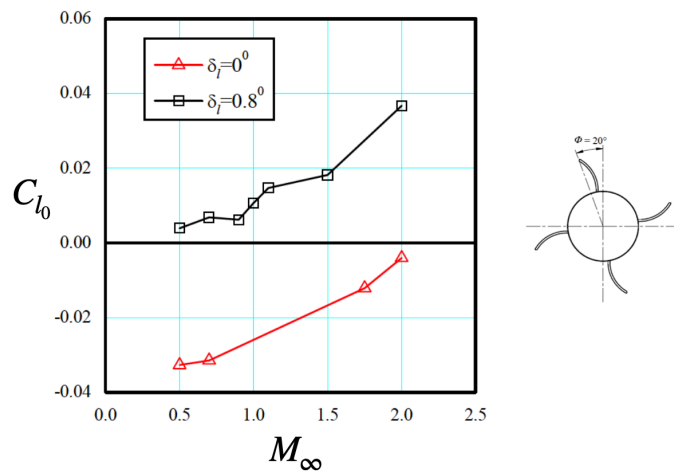

(b) convex side leading into the roll

**Fig. 1 Induced rolling moment coefficient variation with Mach number for two wraparound fin configurations as viewed from behind: (a) with concave side leading, and (b) with convex side leading into the roll (reproduced from Ref. [18]).**

In the following, we will focus on the difference in the sign of the induced rolling moment in the subsonic regime between the two cases illustrated in Fig. 1.

## III. Discrete approximation for curved fin

It is important to recognize that the key parameter that distinguishes the wraparound fin from its planar counterpart is the span-wise curvature. In fact, Ref. [14] did consider the fin radius of curvature as one the parameters in their computational study; however, for the induced rolling moment, they considered only the fin cant angle and fin span-to-chord ratio as the relevant parameters.

In considering the curved wraparound fin, we shall approximate the span-wise curvature by two discrete linear segments, as indicated in Fig. 2 by the straight lines with the filled-circle end points. From Fig. 2, each curved fin appears to be represented by a main wing segment and a deflected wingtip segment. The excluded angle between the two line segments is indicated in Fig. 2; this represents the wingtip segment deflection angle. The precise value of this deflection angle depends, of course, on the choice of the discrete segments, but this is not very significant because our aim is only to obtain a qualitative understanding. Suffice it to say that for most standard wraparound fin geometries, this excluded angle may be expected to be in the mid-acute angle range.

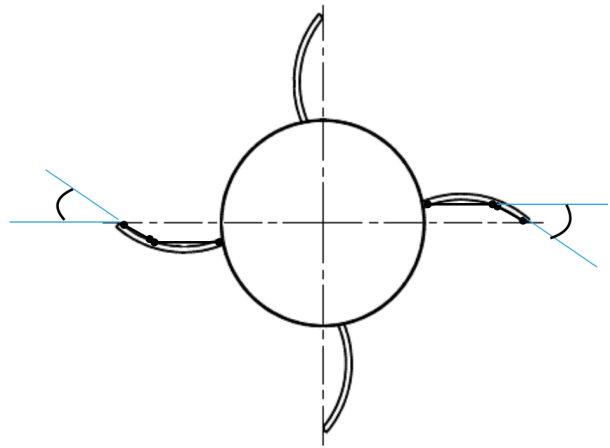

**Fig. 2  Curved fin approximated by two discrete linear segments (viewed from behind).**

Fortuitously, due to the surging interest in airplane wings with articulated wingtips, there is a considerable literature on the aerodynamic properties of lifting surfaces having wingtips with varying deflection. We shall make use of the experimental data reported in Ref. [19] where a flat-plate rectangular wing with aspect ratio of *1.0* having an articulated winglet segment has been tested at low speeds for a Reynolds number of *0.1* million. While this Reynolds number is on the lower side, there is no sign of laminar flow features in the results reported in Ref. [19], so one may be justified in assuming that transition to turbulent flow has occurred and the data in Ref. [19] are applicable for our wraparound fin approximation in the subsonic regime.

The data in Ref. [19] are reported in the following manner. The left wing is held as planar while the right wing has the wingtip deflected with the deflection angle defined as positive for tip down and negative for tip up. If the wingtip deflection results in a loss of lift on the right wing, there will be a positive rolling moment due to the lift imbalance between the two wing halves. On the other hand, if the net rolling moment is approximately zero, or even negative, it would mean that the effect of the right wingtip deflection has been to maintain the lift unchanged (equal to that of the planar wing), or even to augment the lift on the deflected wing half.

The rolling moment data from Ref. [19] are reproduced in Fig. 3. Figure 3(a) shows the rolling moment generated for various wingtip down deflection cases in the experimental arrangement of Ref. [19]. Except for small wingtip deflection angles where there is little to no effect, for all the other deflection cases, the rolling moment is negative. That is, the downward wingtip deflection causes a loss in lift on the wing on its side as compared to that of the planar wing on the other side. This is as expected since the effect of the wingtip deflection is to reduce the effective wing span, which directly affects the lift generated. The same effect should be visible for the wingtip up deflection cases in the data in Fig. 3(b); however, there are several instances here where the loss in lift appears to be lower. In fact, there are data points in Fig. 3(b) with a negative rolling moment, that is, the effect of the wingtip deflection in these cases is to actually augment the lift on its side so that it outweighs the planar wing lift. In particular, consider the data for wing tip deflection of *45* deg (down) and *-45* deg (up), which is a reasonable value for the excluded angle in Fig. 2 for the wraparound fin. Focusing on the lower angle of attack region, Fig. 3(a) shows a consistent loss in lift due to wingtip down deflection at *45* deg, whereas Fig. 3(b) indicates almost no loss in lift for the *45* deg up deflection case. Translated to the WAF approximation as sketched in Fig. 2, this implies that the right-side fin with downward tip deflection has less lift than the one on the left side which has an upward fin tip deflection. Naturally, the resultant rolling moment is positive, as seen in Fig. 1(a). In an identical fashion, for the WAF configuration in Fig. 1(b) rolling with the convex side leading into the roll, one can conclude that the fin curved upwards generates more lift as compared to the other, and hence the resultant rolling moment is negative.

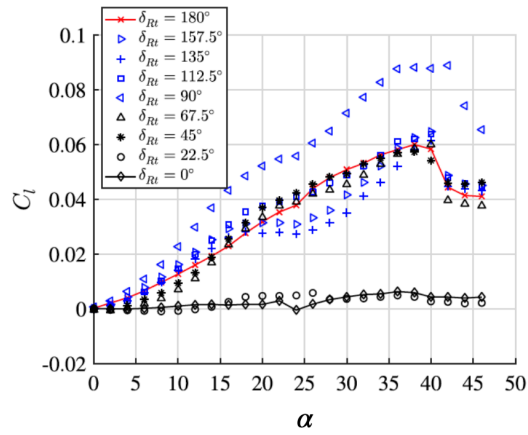

(a) wingtip down deflection

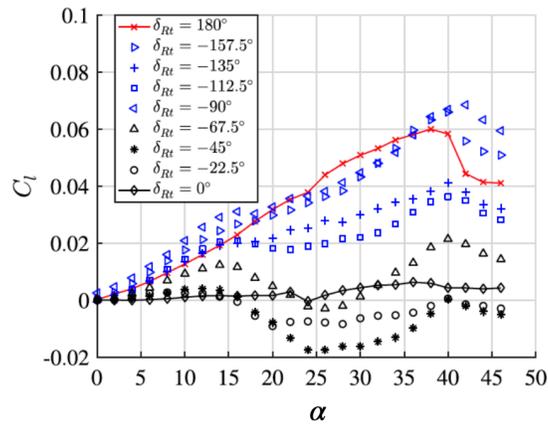

(b) wingtip up deflection

**Fig. 3  Rolling moment coefficient variation with angle of attack for wing with articulated wingtip: (a) for wingtip down deflection, and (b) wingtip up deflection (reproduced from Ref. [19]).**

Thus, it is clear that in case the fins on both sides are up-lifting, then a positive rolling moment is generated for the arrangement in Fig. 1(a) and a negative rolling moment for the configuration in Fig. 1(b). Note, however, that the plots in Fig. 1 are for the rolling moment coefficient at zero free stream angle of attack, where, presumably, the net lift from the oppositely-placed set of fins is approximately zero. So, we need to understand the variation of the induced rolling moment coefficient with angle of attack for small angles of attack.

## IV. Induced rolling moment at small angles of attack

It is clear from the data in Ref. [19] that the induced rolling moment coefficient is almost unchanged for small angles of attack (say, between *-4* and *+4* deg) in either case (that is, with concave side leading into the roll as in Fig. 1(a) or with the convex side leading as in Fig. 1(b)). The same observation was also made in Ref. [20] based on experimental data from fins with various parametric combinations; while the trend over a larger range of angle of attack values was variable depending on the choice of the parameters, all the cases tended to show near-constant values of the rolling moment coefficient in a narrow range near zero angle of attack. We shall try to explain this observation for the fin configuration in Fig. 1(a) using the sketch in Fig. 4; the explanation for the case in Fig. 1(b) is along similar lines and is not repeated here.

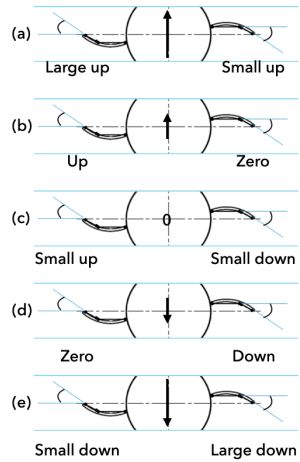

**Fig. 4  Lifting characteristics of a pair of wraparound fins at small free stream angles of attack varying from positive (a,b) through zero (c), and negative (d,e) as viewed from behind.**

Figure 4 shows schematically the lifting characteristics of a pair of oppositely placed wraparound fins at different free stream angles of attack ranging from positive (a), less positive (b), zero (c), slightly negative (d), and more negative (e). The lift on an individual fin in Fig. 4 is indicated below that fin in a qualitative sense. The net lift due to the pair of fins is marked by an arrow in the middle with the direction and length of the vector signifying the sense and magnitude of the lift, respectively. It can be appreciated that the lifting characteristics of an individual fin depends on the local angle of attack, which may be somewhat different from the free stream value. Also, the local angle of attack may differ between the port and starboard fin locations. The data in Fig. 3 refer to the local angle of attack, of course.

Panel (a) in Fig. 4 shows the case discussed in the previous section where the fins on both sides are up-lifting and the difference in their lift yields a positive rolling moment. At a lower value of free stream angle of attack (Panel (b)), the fin with the downward-deflected wingtip may generate zero lift, yet the up-lift on the other fin yields (a smaller value) of positive lift and a positive rolling moment. At zero free stream angle of attack (Panel (c)), the two fins produce nearly equal lift in opposite directions so the net lift is pretty much zero, but the positive rolling moment persists. At a slightly negative free stream angle of attack (Panel (d)), the left fin may show zero lift, but the down lift on the right fin provides a negative lift coefficient but a positive rolling moment still. Finally, Panel (e) is the inverse image of Panel (a) at negative angle of attack; when viewed upside down, the roles of the left and right fins are reversed, but the effect on the rolling moment coefficient is the same. While the net lift does vary with free stream angle of attack from Panel (a) through (e) in the expected manner, the induced rolling moment, which depends on the lift differential between the pair of fins, is almost the same in every case in a qualitative sense, matching the experimental data in Refs. [19,20]. In this manner, it is possible to understand how a pair of wraparound fins may show a variation of net lift with angle of attack but still return an anomalous rolling moment of the same sign, based on the experimental data in Fig. 3 and the geometric representation of the curved fin in Fig. 2.

## V. Mechanism for induced rolling moment

It only remains to decipher the underlying physical mechanism responsible for the generation of the induced rolling moment in wraparound fins as revealed above. From the discussion in Ref. [19], it becomes evident that the difference in the lifting characteristics between the two wingtip deflection cases — the wingtip down case in Fig. 3(a) and the wingtip up case in Fig. 3(b) — is largely due to the position of the wingtip vortex relative to the wing; in particular, their span-wise location and liftoff height. For the WAF configuration discussed in Section III, the location of the wingtip vortices on the two sides is qualitatively depicted in Fig. 5. For the wingtip deflected downwards (the one on the right in Fig. 5), the tip vortex tends to be displaced outboard along the span, whereas the tip vortex on the other fin with the upward-deflected wingtip is shifted inboard. This inboard shift of the vortex places it in a good position to induce additional lift over that fin, and this is the cause of the differential lift between the fins on opposite sides. Thus, the underlying mechanism for the anomalous rolling moment seen in wraparound fin projectiles can be traced to the formation and positioning of the tip vortices at the wingtips of the curved fins.

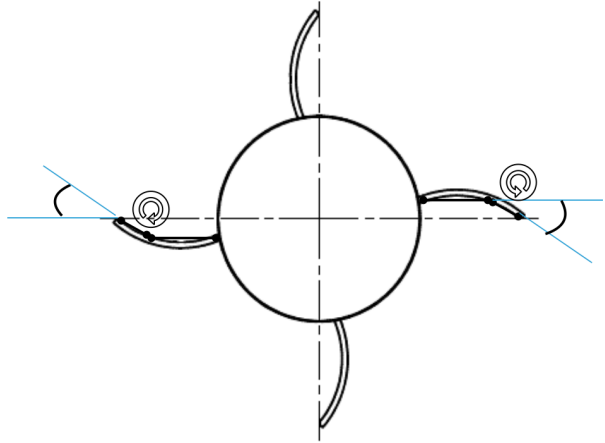

**Fig. 5  Location of the tip vortex on the wraparound fin on either side in the up-lifting case (viewed from behind).**

## V.             Conclusion

A reasonable mechanism is put forward to explain the occurrence of anomalous rolling moments in wraparound fin projectiles. These rolling moments arise due to the differences in the lift induced on a pair of oppositely-placed fins due to the different locations of the wingtip vortex on each side. By modeling the curved fin by a planar segment with another deflected wingtip segment, and using available data in the literature for such wing configurations, the difference in the lifting characteristics between a downward-deflected wingtip and the upward-deflected wingtip on the opposite side can be clearly brought out. Based on these lifting characteristics, the observed sign of the rolling moment coefficient depending on the WAF configuration (that is, having concave surface leading into the roll or the convex surface leading) can be predicted, and the near-constant induced rolling moment at small free stream angles of attack can be explained. Hopefully, future studies will be able to experimentally or numerically image the vortex structure over the curved WAF fins to verify the mechanism proposed in this work. Work also remains to be done to understand the mechanisms behind the induced rolling moment at supersonic Mach numbers and in the transition Mach regime.

doi: 10.2514/3.57135

[2] Catani, U., Bertin, J. J., De Amicis, R., Masullo, S., and Bouslog, S.A., "Aerodynamic Characteristics for a Slender Missile with Wraparound Fins," *Journal of Spacecraft and Rockets*, Vol. 20, No. 2, 1983, pp. 122-128.

doi: 10.2514/3.28367

[3] Lorenz, R. D., *Spinning Flight: Dynamics of Frisbees, Boomerangs, Samaras, and Skipping Stones*, Springer, New York, 2006, p. 79, Fig. 4.7.

[4] Murphy, C. H., "Some Special Cases of Spin-yaw Lock-in," *Journal of Guidance, Control, and Dynamics*, Vol. 12, No. 6, 1989, pp. 771-776.

doi: 10.2514/3.20480

[5] Ananthkrishnan, N., and Raisinghani, S. C., "Steady and Quasisteady Resonant Lock-in of Finned Projectiles," *Journal of Spacecraft and Rockets*, Vol. 29, No. 5, 1992, pp. 692-696.

doi: 10.2514/3.11512

[6] Sharma, A., and Ananthkrishnan, N., "Passage through Resonance of Rolling Finned Projectiles with Center of Mass Offset," *Journal of Sound and Vibration*, Vol. 239, No. 1, 2001, pp. 1-17.

doi: 10.1006/jsvi.2000.3114

[7] Liaño, G., and Morote, J., "Roll-Rate Stability Limits of Unguided Rockets with Wraparound Fins," *Journal of Spacecraft and Rockets*, Vol. 43, No. 4, 2006, pp. 757-761.

doi: 10.2514/1.17775

[8] Mikhail, A.G., "Roll Damping for Projectiles including Wraparound, Offset, and Arbitrary Number of Fins," *Journal of Spacecraft and Rockets*, Vol. 32, No. 6, 1995, pp. 929-937.

doi: 10.2514/3.26711

[9] Kim, Y. H., and Winchenbach, G. L., "Roll Motion of a Wraparound Fin Configuration at Subsonic and Transonic Mach Numbers," *Journal of Guidance, Control, and Dynamics*, Vol. 9, No. 2, 1986, pp. 253-255.

doi: 10.2514/3.20100

[10] McIntyre, T. C., Bowersox, R. D. W,. And Goss, L. P., "Effects of Mach Number on Supersonic Wraparound Fin Aerodynamics," *Journal of Spacecraft and Rockets*, Vol. 35, No. 6, 1998, pp. 742-748.

doi: 10.2514/2.3410

[11] Paek, S. K., Park, T. S., Bae, J. S., Lee, I., And Kwon, J. H., "Computation of Roll Moment for Projectile with Wraparound Fins using Euler Equations," *Journal of Spacecraft and Rockets*, Vol. 36, No. 1, 1999, pp. 53-58.

doi: 10.2514/2.3432

[12] Edge, H. L., "Computation of the Roll Moment for a Projectile with Wraparound Fins," *Journal of Spacecraft and Rockets*, Vol. 31, No. 4, 1994, pp. 615-620.

doi: 10.2514/3.26486

12 of 12

12 of 12